\documentclass[prx,aps,twocolumn,showpacs,amsmath,amssymb,prb,footinbib,floatfix,superscriptaddress]{revtex4-1}
\usepackage{graphicx}
\usepackage{savesym}
\usepackage{amsfonts}
\usepackage{bm}
\usepackage{color}
\usepackage{hyperref}
\usepackage{subfigure}
\usepackage{wasysym}
\usepackage{upgreek}
\usepackage{float}
\usepackage{dsfont}
\usepackage{mathtools}

\usepackage[utf8]{inputenc}
\usepackage{pgfplots}
\pgfplotsset{compat=newest}
\usepgfplotslibrary{groupplots}
\usepgfplotslibrary{dateplot}

\newcommand{\lo}[1]{{\color{red} #1}}

\begin{document}

\title{Probabilistic Simulation of Quantum Circuits with the Transformer} 

\author{Juan Carrasquilla}
\affiliation{Vector Institute, MaRS Centre, Toronto, Ontario, M5G 1M1, Canada}

\author{Di Luo}
\affiliation{Institute for Condensed Matter Theory and IQUIST and Department of Physics,
University of Illinois at Urbana-Champaign, IL 61801, USA}

\author{Felipe P\'erez}
\affiliation{Layer6 AI, MaRS Centre, Toronto, Ontario, M5G 1M1, Canada}

\author{Ashley Milsted}
\affiliation{Perimeter Institute for Theoretical Physics, 31 Caroline Street North, Waterloo, ON N2L 2Y5, Canada}

\author{Bryan K. Clark}
\affiliation{Institute for Condensed Matter Theory and IQUIST and Department of Physics,
University of Illinois at Urbana-Champaign, IL 61801, USA}

\author{Maksims Volkovs}
\affiliation{Layer6 AI, MaRS Centre, Toronto, Ontario, M5G 1M1, Canada}

\author{Leandro Aolita}
\affiliation{ Instituto de F\'isica, Universidade Federal do Rio de Janeiro, Caixa Postal 68528,
Rio de Janeiro, RJ 21941-972, Brazil}

\date{\today}

\begin{abstract}
The fundamental question of how to best simulate quantum systems using 
conventional computational resources lies at the forefront of condensed 
matter and quantum computation.  It impacts both our understanding of 
quantum materials and our ability to emulate quantum circuits.  Here 
we present an exact formulation of quantum dynamics via factorized 
generalized measurements which maps quantum states to probability 
distributions with the advantage that local unitary dynamics and quantum 
channels map to local quasi-stochastic matrices. This representation 
provides a general framework for using state-of-the-art probabilistic 
models in machine learning for the simulation of quantum many-body dynamics.  
Using this framework, we have developed a practical algorithm to simulate 
quantum circuits with the Transformer, a powerful ansatz responsible for 
the most recent breakthroughs in natural language processing. We 
demonstrate our approach by simulating circuits which build GHZ and linear 
graph states of up to 60 qubits, as well as a variational quantum eigensolver 
circuit for preparing the ground state of the transverse field Ising model 
on six qubits.  Our methodology constitutes a modern machine learning approach 
to the simulation of quantum physics with applicability both to quantum 
circuits as well as other quantum many-body systems. 
\end{abstract} 

\maketitle


In his celebrated keynote address at the California Institute of Technology in
May 1981, Feynman introduced the idea of a computer that could act as a quantum
mechanical simulator,\cite{feynman1982} which has inspired the field of quantum
computing since its inception. In his keynote, Feynman also intriguingly asked 
``can quantum systems be probabilistically simulated by classical
computer?'', which he answered negatively observing that a probabilistic simulation
is unfeasible since the description of both the quantum state and its evolution
necessarily involves non-positive quasi-probabilities.

These simple but powerful observations provide a glimpse into the boundary of
what behaviour can be considered classical or quantum, since classical systems are
fully characterized by traditional probability distributions evolving according
to non-negative stochastic matrices.\cite{domb2000phase} From a computational 
viewpoint, these observations are also fundamentally linked to the notion of 
quantum speed-up in quantum computing, where it is anticipated that quantum 
computers will display potential speed-ups over their classical counterparts 
at the onset of negative values in the quasi-probabilities associated with the 
description and evolution of their quantum 
states.\cite{ferrie2011,veitch2012,raussendorf2019} This is in part due to 
the negative values in the quasi-probability distributions 
precluding an efficient simulation of quantum dynamics through Monte Carlo 
techniques due to the debilitating sign problem induced by the negative values.

Although the presence of negative quasi-probabilities is often linked to intrinsically
quantum phenomena with no classical counterpart like entanglement and quantum
interference, a purely probabilistic representation of the quantum state
is possible.\cite{fuchs2011a,chruscinski2015,vandewetering2018,carrasquilla2019a,kiktenko2019}
While in the standard formulation of quantum mechanics a quantum state is represented 
by a density operator, a quantum state can also be completely specified by the outcome 
probability of a physical measurement, provided that the measurement probes enough 
information about the quantum state. This notion is made precise through two fundamental 
concepts in quantum theory: the so-called Born rule, which is the theoretical principle of 
quantum physics linking quantum theory and experiment, and the concept of 
informationally complete (IC) measurements, which are described by positive-operator 
valued measures (POVMs). Whereas POVMs describe the most general 
type of measurements allowed by quantum theory going beyond the notion of projective 
measurements,~\cite{Nielsen} informational completeness means that the 
outcome statistics of such a measurement specifies the quantum state unambiguously. 

Recently, neural probabilistic models (NPM) that are commonly used in language 
modelling and translation have been used in the context of quantum state reconstruction.\cite{carrasquilla2019a}
Such a strategy resulted in an accurate probabilistic representation of families of 
prototypical states in quantum information as well as complex ground states of 
one- and two-dimensional local Hamiltonians describing large many-body systems 
relevant to condensed matter, cold atomic systems, and quantum simulators.\cite{carrasquilla2019a} 
Although an exact simulation of quantum dynamics using probability 
remains, in general, unfeasible, it is natural to ask whether the power and 
scalability of the neural probabilistic representation of the quantum state in Ref.~\onlinecite{carrasquilla2019a}, 
can be harnessed in the classical simulation of quantum systems, e.g. in 
challenging scenarios like the study of real- and imaginary-time dynamics of 
many-body systems, or in the classical simulation of quantum circuits and quantum 
channels. 

Inspired by Feynman's notion of a probabilistic simulation of quantum systems, 
we employ a probabilistic representation of the quantum state and 
combine it with state-of-the-art neural language modelling architectures 
to approximate the statistics of the measurement outcome of states produced by
quantum circuits.  The
near-term availability of quantum computers with the potential 
to perform computational tasks beyond the capabilities of the most 
powerful classical computers\cite{supremacy} has given rise to a new generation 
of exact and approximate classical simulation techniques\cite{deraedt2007,markov2008,smelyanskiy2016,pednault2017,
chen2018ClassicalSimulation,markov2018,li2018} for quantum circuits, including 
a neural network approach.\cite{jonsson2018} We test our ideas by considering 
quantum circuits which prepare prototypical states in quantum information. In 
particular we consider the GHZ state, linear graph state, and the variational ground state of 
the transverse field Ising model (TFIM). Through numerical experiments, we show 
that our strategy produces accurate results for the target states of up to 
60 qubits, which opens up a new probabilistic avenue for simulation of quantum 
circuits, as well as quantum channels and quantum dynamics more broadly. 

\begin{figure*}
\centering
\includegraphics[width=0.85\textwidth]{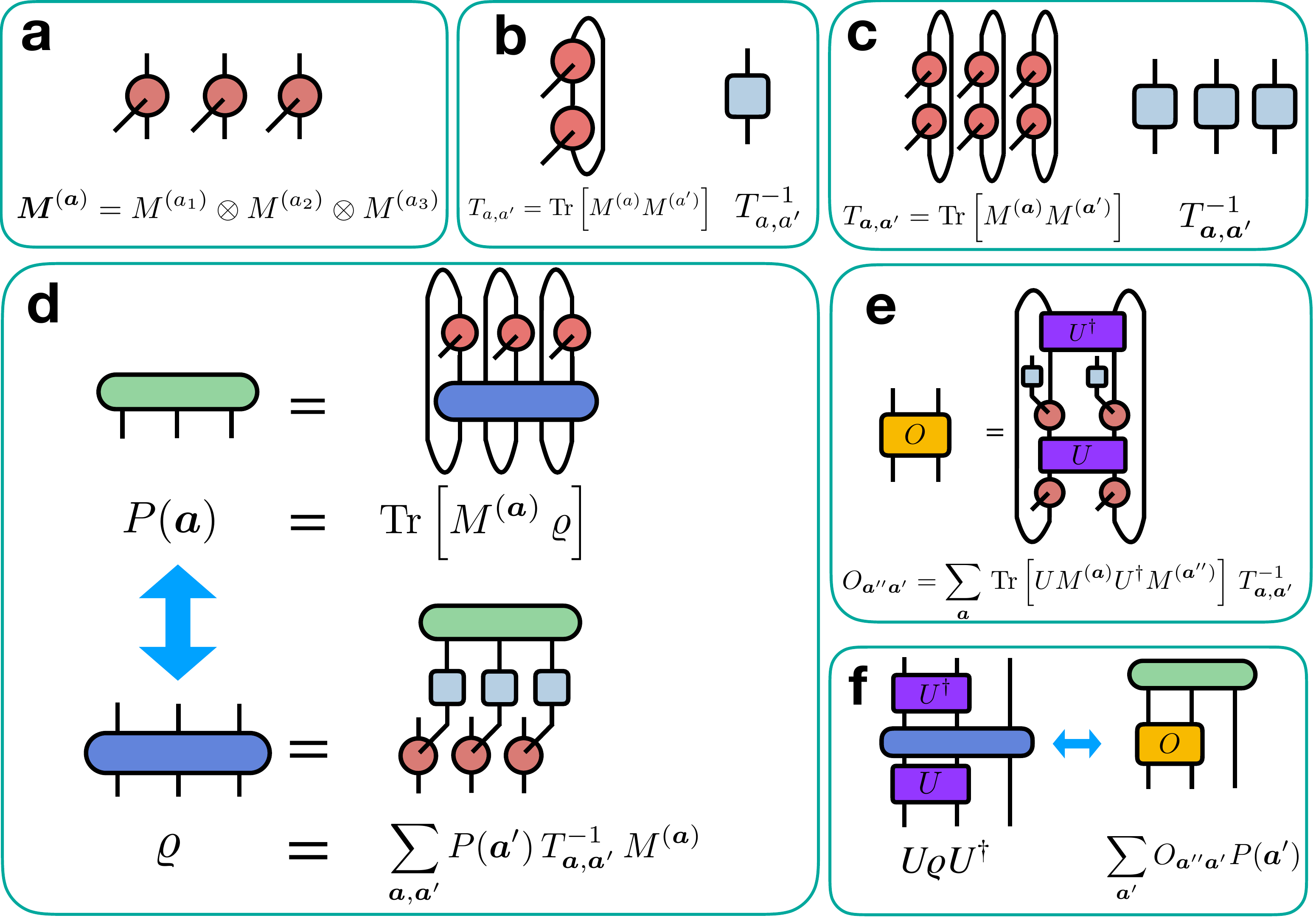}\newline
\caption{Tensor-network representation of the mapping between quantum states (gates) 
and probability distributions (quasi-stochastic matrices) used in this work. (a) 
$N$-qubit measurement $\boldsymbol{M}=\big\{M^{(a_1)}\otimes M^{(a_2)}\otimes 
\hdots M^{(a_N)}\big\}_{a_1, \hdots a_N}$ made from ($N=3$) single qubit measurements 
 $\boldsymbol{M}=\{M^{(a)}\}_a$ (red). Vertical indices in the
red tensors act on the physical degrees of freedom (qubits) while the horizontal 
index labels the measurement outcome $a$. (b) The overlap matrix 
$T_{\boldsymbol{a_1},\boldsymbol{a_1}'}$ and its inverse $T^{-1}$(light blue) 
(c) multi-qubit version of b. (d) The Born rule relates the probability 
$P(\boldsymbol{a})$ (green; indices encode the different measurement 
outcomes on each qubit) to the quantum state $\varrho$ (blue). (e) Unitary 
gates (purple) map to quasi-stochastic matrices (yellow). (f) Application of a 
unitary matrix to a density matrix corresponds to applying a quasi-stochastic 
matrix to $P$.
\label{fig:TNrep}}
\end{figure*}

\noindent \textbf{Formalism}: We focus on physical systems composed of $N$ 
qubits whose quantum state, traditionally represented by a density matrix 
$\rho$, will be uniquely specified by the measurement statistics of an 
informationally complete POVM (IC-POVM). To build an IC-POVM for $N$ qubits, 
we first consider an $m$-outcome single-qubit IC-POVM defined by a collection  
$\boldsymbol{M}=\{M^{(a)}\}_{a \in \{1..m\}}$, of positive semi-definite operators 
$M^{(a)}\geq 0$, each one labeled by a measurement outcome $a=0,1,..,m-1$ 
[see Fig.\ref{fig:TNrep}(a) where we describe our representation through the lens 
of tensor networks and its graphical notation.\cite{penrose1971}]  Following 
Ref.~\onlinecite{carrasquilla2019a}, we construct $N$-qubit measurements 
as tensor products of the single-qubit IC-POVM elements 
$\boldsymbol{M}=\big\{M^{(a_1)}\otimes M^{(a_2)}\otimes 
\hdots M^{(a_N)}\big\}_{a_1, \hdots a_N \in \{1..m\}^N}$, 
as graphically depicted in Fig.\ref{fig:TNrep}(a). Due to the 
simplicity of its implementation with currently available gate-based quantum 
computers, in our numerical simulations we consider the 4-Pauli IC-POVM measurement 
described in Ref.\onlinecite{carrasquilla2019a}. Born's rule predicts that the 
probability distribution $\boldsymbol{P}=\{P(\boldsymbol{a})\}_{\boldsymbol{a}=(a_1,a_2,...,a_N)}$ 
over measurement outcomes $\boldsymbol{a}$ on a quantum state $\varrho$ is given
by $P(\boldsymbol{a})=\text{Tr}\left[ M^{(\boldsymbol{a})}\, \varrho\right]$, as 
graphically explained in Fig.\ref{fig:TNrep}(d). Note that a quantum state is 
specified by $m^{N}$ probabilities. Due to the factorized nature of the IC-POVM, 
a product state $\prod_i \otimes |\Psi_i\rangle$ takes the form of a product 
distribution over statistically independent sets of variables 
$P(\boldsymbol{a})=P(a_1)P(a_2)\cdots P(a_N)$  where 
$P(a_i)=Tr[M^{(a_i)} |\Psi_i\rangle\langle \Psi_i |]$.  Provided that 
the measurement is informationally complete,  the density 
matrix can be inferred from the statistics of the measurement outcome as
\begin{equation}\label{Eq:rho}
\begin{split}
    \varrho &= \sum_{\boldsymbol{a},\boldsymbol{a}'} P(\boldsymbol{a}')\,
T^{-1}_{\boldsymbol{a},\boldsymbol{a}'}\,M^{(\boldsymbol{a})}, 
\end{split}
\end{equation}
where $T$ represents the overlap matrix given by 
$T_{\boldsymbol{a},\boldsymbol{a}'} = \text{Tr}\left[  M^{(\boldsymbol{a})} M^{(\boldsymbol{a}')} \right]$. 
See Fig.\ref{fig:TNrep}(d) for a graphical representation of elements in Eq.\ref{Eq:rho}.

To study quantum circuits, we first have to translate the action of a quantum gate 
on the density matrix to the IC-POVM representation. The former corresponds to a unitary 
transformation, i.e. $\varrho_{U}= U \varrho U^{\dag}$. 
If the initial quantum state is prescribed in terms of the outcome statistics
of an IC-POVM $\boldsymbol{P}$, we can track its evolution directly in the probabilistic 
representation:
 
\begin{equation} \label{eq:stoevol}
P_{U}\left(\boldsymbol{a}'' \right) = \text{Tr}\left[U \varrho U^{\dag} M^{(\boldsymbol{a}'')} \right] = \sum_{\boldsymbol{a}'} O_{\boldsymbol{a}''\boldsymbol{a}'} P(\boldsymbol{a}')
\end{equation}
where 
\begin{equation}\label{SWStoch}
O_{\boldsymbol{a}''\boldsymbol{a}'}=\sum_{\boldsymbol{a}} \,\text{Tr}\left[U M^{(\boldsymbol{a})} U^{\dag} M^{(\boldsymbol{a}'')} \right] \,T^{-1}_{\boldsymbol{a},\boldsymbol{a}'}
\end{equation}
is a \emph{somewhat} stochastic matrix since the values in 
each column add up to 1 but its entries can be positive or 
negative.\cite{curgus2007,curgus2015,chruscinski2015,vandewetering2018}
Somewhat stochastic matrices are also known as pseudo-stochastic 
or quasi-stochastic matrices.\cite{chruscinski2015,vandewetering2018}
We note that the evolution described in Eq.~\ref{eq:stoevol} leads to a 
formulation of quantum mechanics equivalent to, e.g. Heisenberg’s matrix 
mechanics, including the description of open quantum systems, quantum channels, and
measurements of other POVMs (See Appendix). 

Here we emphasize that Eq.~\ref{eq:stoevol} resembles the standard rule for 
stochastic evolution commonly used to describe the transitions in a 
Markov chain, where the traditional stochastic (or Markov) matrix has 
been replaced with a  quasi-stochastic matrix. Despite the resemblance, 
a generic classical Markov Chain Monte Carlo simulation of quantum 
evolution in the probabilistic factorized POVM language remains 
unfeasible due to the numerical sign problem arising from the negative 
entries of the quasi-stochastic matrix describing the process.

Due to the factorized nature of the IC-POVM, if a unitary matrix 
or a quantum channel acts nontrivially on only $k$ qubits of the quantum 
system, the quasi-stochastic matrix $O_{\boldsymbol{a}''\boldsymbol{a}'}$ 
acts only on the measurement outcomes of those $k$ qubits too. For 
example, a two-qubit unitary gate acting on qubits $i$ and $j$ is 
represented by a $m^2\times m^2$ quasi-stochastic matrix acting on 
outcomes $a_i$ and $a_j$. The relation between the local 
quasi-stochastic matrices and the local unitary gates, 
as well as their action on a quantum  state are graphically depicted 
in  Fig.\ref{fig:TNrep}(e-f) using tensor diagrams. Furthermore, 
the locality of $O_{\boldsymbol{a}''\boldsymbol{a}'}$ implies that 
traditional  quantum circuit diagrams\cite{Nielsen} translate 
into probabilistic circuits that look exactly the same as their traditional 
counterparts.

A quantum circuit is a generalization of the circuit model of classical 
computation where a product state is evolved through a series 
of unitary gates, $U^{(1)}, U^{(2)},\cdots, U^{(r)}$, 
each of which acts nontrivially on a constant number $k$ qubits. Thus, 
the quantum circuit in the IC-POVM representation becomes a computation where
an initial probability distribution $P(\boldsymbol{a})=P(a_1)P(a_2)\cdots P(a_N)$   
of statistically independent sets of variables $\boldsymbol{a}$ which is 
evolved through a series of local quasi-stochastic matrices of the form 
depicted in Fig.\ref{fig:TNrep}(e). 
The measurement statistics after unitary evolution through the 
first gate $U^{(1)}$ is given by $\boldsymbol{P}_{1}=O^{(1)}\boldsymbol{P}_{0}$, 
and the application of each subsequent gate $U^{(i)}$ defines a 
series of intermediate probability distributions  
$\boldsymbol{P}_{i}=O^{(i)} O^{(i-1)}\cdots O^{(2)} O^{(1)}\boldsymbol{P}_{0}$ 
with $i=1,\cdots,r$. 

\begin{figure}
\centering
\includegraphics[width=0.75\linewidth]{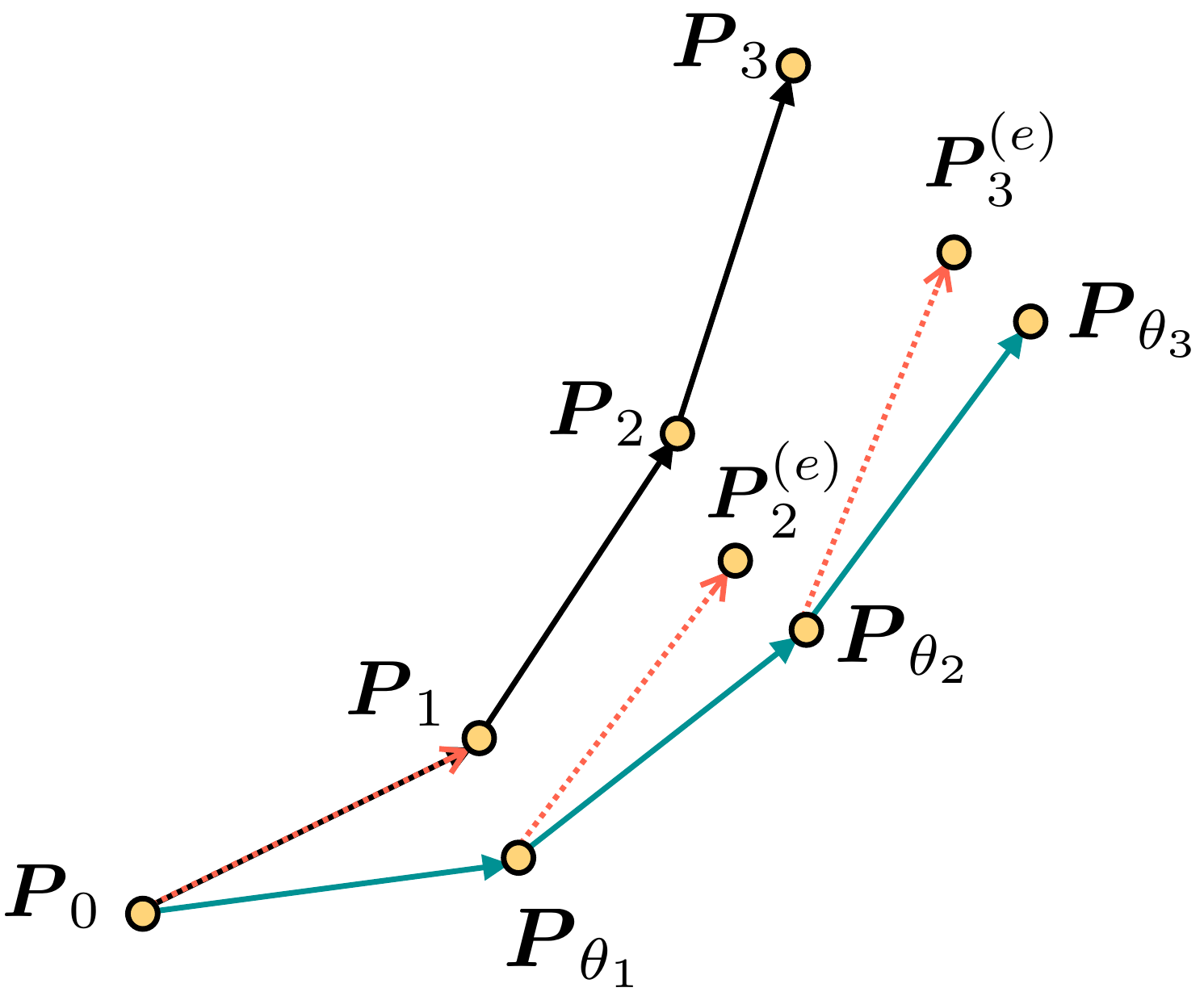}\newline
\caption{ Schematics of the different distributions involved in the training procedure
for a circuit with 3 gates.
Black arrows indicate the exact trajectory while green arrows represent the trajectory 
of the optimized models $\boldsymbol{P}_{\theta_i}$. The red dotted arrows
point to the exactly evolved distributions $\boldsymbol{P}^{(e)}_{i+1}$ 
after the application of each  gate $O^{(i)}$ on the trained model 
$\boldsymbol{P}_{\theta_{i}}$. Note that if the resulting KL divergence after training
$\text{KL}\left(\boldsymbol{P}^{(e)}_{i+1}|| \boldsymbol{P}_{\theta_{i+1}}\right)=0$ 
for every gate in the circuit, then the black trajectory coincides with the green 
one and the procedure becomes exact.}
\label{fig:process}
\end{figure}

\noindent \textbf{Algorithm}: 
The strategy to learn the output distribution $\boldsymbol{P}_{r}$ consists in 
constructing models $\boldsymbol{P}_{\theta_{i}}=\{P(\boldsymbol{a};
\theta_{i})\}_{\boldsymbol{a}=(a_1,a_2,...,a_N)}$
based on a rich family of probability distributions 
$\boldsymbol{P}(\boldsymbol{a}; \theta)$. These are expressed in terms of 
a highly expressive neural network with parameters $\theta$ 
so that $\boldsymbol{P}_{\theta_{i}} \approx \boldsymbol{P_{i}}$. At 
each time step $i$, we assume that an accurate neural approximation has 
been reached $\boldsymbol{P}_{\theta_i} \approx \boldsymbol{P}_{i}$, and 
consider the exactly evolved distribution  
$\boldsymbol{P}^{(e)}_{i+1} \equiv O^{(i+1)}\boldsymbol{P}_{\theta_i}$. While the 
representation of the quantum state at step $i$ isn't exact, if 
$\boldsymbol{P}_{\theta_i}$ is sufficiently accurate the expectation is 
that the distribution $\boldsymbol{P}_{\theta_{i+1}}\approx \boldsymbol{P}^{(e)}_{i+1}$. 
See Fig.~\ref{fig:process} for a depiction of the distributions involved 
during the simulation.

To train the model $\boldsymbol{P}_{\theta_i}$ given a gate $i+1$, we adopt 
a variational approach and select the parameters $\theta_{i+1}$ such 
that the Kullback-Liebler (KL) divergence 
between $\boldsymbol{P}^{(e)}_{i+1}$ and $\boldsymbol{P}_{\theta_{i+1}}$,
\begin{equation}
\label{eq:KL}     \text{KL}\left(\boldsymbol{P}^{(e)}_{i+1}|| \boldsymbol{P}_{\theta_{i+1}} \right) \\
    =   -\sum_{\boldsymbol{a}}P^{(e)}_{i+1}(\boldsymbol{a}) \log\left( \frac{P_{\theta_{i+1}}(\boldsymbol{a})}{P^{(e)}_{i+1}(\boldsymbol{a})}\right) 
\end{equation}
is minimized. This objective function is reasonable as
$\text{KL}\left(\boldsymbol{P}^{(e)}_{i+1}|| \boldsymbol{P}_{\theta_{i+1}}\right)\geq 0 $, 
with the equality being satisfied when $\boldsymbol{P}_{\theta_{i+1}}= \boldsymbol{P}^{(e)}_{i+1}$.   
To minimize the KL divergence, we will apply a variant of gradient descent 
(i.e. ADAM\cite{Adam}) where we repeatedly update the parameters of 
$\theta_{i+1}$ by taking steps in the direction of the gradient of 
Eq.~\ref{eq:KL}. This gradient, assuming that the model is normalized, 
can be written as 

\begin{subequations}
\begin{align}\label{eq:gradient}
&\nabla_{\theta_{i+1}} \text{KL}(\boldsymbol{P}^{(e)}_{i+1}|| \boldsymbol{P}_{\theta_{i+1}}) \\
&= -\sum_{\boldsymbol{a}}P^{(e)}_{i+1}(\boldsymbol{a}) \nabla_{\theta_{i+1}}  \log\left(P_{\theta_{i+1}}(\boldsymbol{a})\right) \label{b}\\
&= -\mathbb{E}_{\boldsymbol{a}\sim P_{\theta_{i+1}}(\boldsymbol{a})} \left[  \frac{P^{(e)}_{i+1}(\boldsymbol{a})}{P_{\theta_{i+1}}(\boldsymbol{a})} \nabla_{\theta_{i+1}}  \log\left(P_{\theta_{i+1}}(\boldsymbol{a})\right)\right] \\
&= -\mathbb{E}_{\boldsymbol{a}\sim P_{\theta_{i+1}}(\boldsymbol{a})}\left[ \left( \frac{P^{(e)}_{i+1}(\boldsymbol{a})}{P_{\theta_{i+1}}(\boldsymbol{a})} - k \right) \nabla_{\theta_{i+1}}  \log\left(P_{\theta_{i+1}}(\boldsymbol{a})\right)\right].
\end{align}
\end{subequations}
Here, $k=\mathbb{E}_{\boldsymbol{a}\sim P_{\theta_{i+1}}(\boldsymbol{a})} \left[\frac{P^{(e)}_{i+1}(\boldsymbol{a})}{P_{\theta_{i+1}}(\boldsymbol{a})} \right]$.   
Note that $\mathbb{E}_{\boldsymbol{a}\sim P_{\theta_{i+1}}(\boldsymbol{a})}   \nabla_{\theta_{i+1}}  \log\left(P_{\theta_{i+1}}(\boldsymbol{a})\right) =0$, which justifies the equality in Eq.~5d.
To estimate the gradient in Eq.~\ref{eq:gradient}, we generate a mini-batch 
of $N_s$ samples of $\boldsymbol{a}$ sampled from $P_{\theta_{i+1}}(\boldsymbol{a})$ 
and average over these samples both to compute the value of $k$ as well as Eq.~5d.  
While computing Eq.~5d or Eq.~5c both evaluate the gradient, Eq.~5d has 
significantly better variance. In the limit where $P_{i+1}^{(e)}(\textbf{a})$ 
approaches $P_{\theta_{i+1}}(\textbf{a})$ (i.e. ideally towards the end of 
the optimization of gate $i+1$), the variance of this 
gradient goes to zero.  This is known as the zero variance 
principle.~\cite{zero_variance} $P^{(e)}_{i+1}(\boldsymbol{a})=
\sum_{\boldsymbol{a}'} O^{(i+1)}_{\boldsymbol{a} \boldsymbol{a}'}P_{\theta_i}(\boldsymbol{a}')$ 
is evaluated by  explicitly summing over the sub-string of outcomes in 
$\boldsymbol{a}'$ on which $O^{(i+1)}$ acts (with the other outcomes in 
the string fixed). 

By construction, the unitary matrices and their corresponding 
quasi-stochastic matrices considered here are $k$-local, which 
means that the calculation of the gradient estimator in 
Eq.~\ref{eq:gradient} is efficient.

\begin{figure}
\centering
\includegraphics[width=0.5\linewidth]{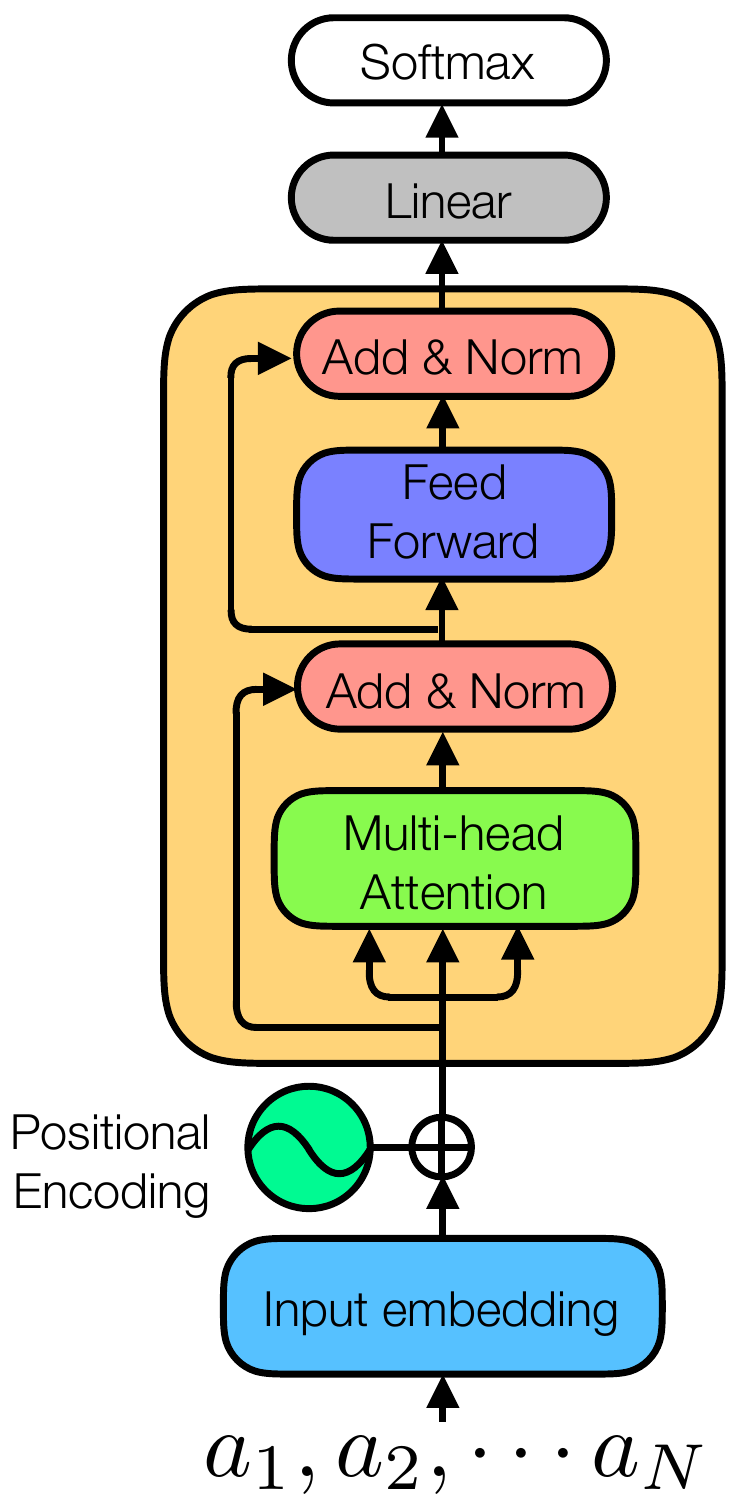}\newline
\caption{ Schematic representation of the Transformer model. Lines with arrowheads denote incoming arrays from the output of one node to the inputs of others. The Transformer architecture
starts with an input measurement $\boldsymbol{a}$.  First, a high-dimensional 
linear embedding $\hat{A}$ of the input measurement is computed. This is followed 
by addition of positional encoding vectors to the input embeddings $\hat{A}$. 
The multi-head attention mechanism is applied to the modified embedding,
followed by a residual connection\cite{he2015a} and layer normalization.\cite{ba2016}
A position-wise feed-forward is then applied the outcome of the previous layer 
again followed by a residual connection and layer normalization. The output of 
the last layer is processed through a linear layer followed by a softmax 
which returns the conditional probabilities $P_{\theta}(a_{k+1}|a_{1}, \hdots, a_k)$.}
\label{fig:transformer}
\end{figure}

\noindent \textbf{Transformer}: For simplicity, in order to model $\boldsymbol{P}_i$, 
we restrict to models $P_{\theta_{i}}(\boldsymbol{a})$ with a 
tractable density and that allow for exact sampling. While other models
such as the variational autoencoder\cite{luchnikov2019} can represent the 
quantum state probabilistically, having both a tractable density and exact sampling 
significantly simplifies the calculation of the quantities involved in the gradient 
estimation. The exact sampling avoids expensive Markov Chain Monte Carlo which 
would otherwise be required to obtain the samples for the gradient estimator. 
Specifically, we consider prototypical autoregressive models commonly used in 
neural machine translation and language modelling based on Transformer encoder
blocks.\cite{AttentionIsAllYouNeed} This neural architecture models a 
probability distribution $P_{\theta_{i+1}}(\boldsymbol{a})$  
through the conditional $P_{\theta}(a_{k+1}|a_{1}, \hdots, a_k)$. 
Note that we can recover $\boldsymbol{P}$ via the chain rule
\[P_{\theta}(a_1, \ldots, a_N) = \prod_{k=1}^{N} P_{\theta}(a_k|a_{<k}),\]
which we heavily use in our simulations. 

The Transformer architecture is constructed using the elements depicted 
in Fig.~\ref{fig:transformer}. The first and most important element is the 
self-attention mechanism. Self-attention takes an embedding of the measurement 
outcome $\boldsymbol{a}$, and computes an auto-correlation matrix 
where the different measurement outcomes across the different qubits 
form the columns and rows. The embedding is simply a linear transformation
on the original input $\boldsymbol{a}$. The self-attention and its correlation
matrix are useful to introduce correlations between qubits separated at 
any distance in the quantum system. This is analogous to a two-body 
Jastrow factor\cite{becca2017} which induces pairwise long-distance 
correlations between the bare degrees of freedom (i.e. spins, qubits, electrons) 
in a wavefunction. In contrast to traditional sequence models based on 
recurrent neural networks, which tend to suppress correlations beyond a 
certain length $\xi$, the self-attention networks are suitable to model 
systems exhibiting power-law correlations present in natural sequences as well as 
physical systems exhibiting (classical or quantum) critical behaviour.\cite{shen2019} 

More precisely, the attention mechanism can be described as a map between 
a ``query'' array $Q$, a ``key'' array $K$  and ``value'' $V$, to an 
output vector. The the query, keys, values, are linear transformations 
of the input vectors, e.g. $K=\hat{A} W^{(K)}$, where 
$\hat{A} \in \mathbb{R}^{N \times d_\text{model}} $ is a 
$d_\text{model}$-dimensional embedding of the measurements outcome 
at the different qubits $j=1,\dots, N$, and $W^{(K)}$ $\in 
\mathbb{R}^{d_{\text{model}} \times d_k}$ is a parameter of the model. 
Analogously, the values and queries are calculated as a parametrized 
linear transformation on the embedding $\hat{A}$.  

The specific type of attention mechanism the Transformer uses is the 
so-called scaled dot-product attention. The input consists of queries 
and keys of dimension $d_k$, and values of dimension $d_v$, and the 
output is computed as 
\begin{equation}
    \text{Attention}\left(Q,K,V\right)= \text{softmax}\left( \frac{Q K^{\text{T}}}{\sqrt{d_k}}\right) V,
\end{equation}
where the softmax function acting on a vector results in 
$\text{softmax}(x_i)=\frac{e^{x_i}}{\sum_{j}e^{x_j}}$. The argument 
of the softmax is $\hat{A} W^{(Q)}   {W^{(K)}}^{\text{T}} \hat{A}^{\text{T}}$, 
which induces pairwise, all-to-all correlations between the qubits in the 
system, thus resembling a Jastrow factor with parameters 
$W^{(Q)} {W^{(K)}}^{\text{T}}$.  

As in Ref.~\onlinecite{AttentionIsAllYouNeed}, we use a multi-head 
attention mechanism where instead of computing a single attention 
function, we linearly project the queries, keys and values $h$ times 
with different, learned linear projections to $d_k$, $d_k$ and $d_v$ 
dimensions. Each of these projections are then followed by the attention 
function in parallel, producing $d_v$-dimensional output values. 
These are concatenated and projected. The output of the multi-head 
attention is 
\begin{equation}
\begin{split}
\text{Multi-Head}\left(Q,K,V\right)=
    \\
     \text{Concat}\left( \text{head}_1,\dots,\text{head}_h\right)W^{(0)},
    \end{split}
\end{equation}
where $\text{head}_i=\text{Attention}\left(Q_i,K_i,V_i\right)$,
$K_i=\hat{A} W_i^{(K)}$, $Q_i=\hat{A} W_i^{(Q)}$, and $V_i=\hat{A} W_i^{(V)}$. 
Here, $W_i^{(K)} \in \mathbb{R}^{d_{\text{model}} \times d_k}$, 
$W_i^{(Q)} \in \mathbb{R}^{d_{\text{model}} \times d_k}$, and 
$W_i^{(V)} \in \mathbb{R}^{d_{\text{model}} \times d_v}$. In our work 
we use  $h = 8$ attention heads, and $d_k = d_v = d_{\text{model}}/h$ 
with $d_{\text{model}} = 16$  or 32.
 
Additionally, the Transformer features a position-wise feed-forward 
network, which is a fully connected feed-forward network applied to 
each position separately and identically. This layer consists of two 
linear transformations with a ReLU\cite{Goodfellow-et-al-2016} 
activation in between. 
 
Each sub-layer (i.e. the self-attention and the position-wise 
feed-forward network) has a residual connection around it, and 
is followed by a layer-normalization step. That is, the output 
of each sub-layer is $\text{LayerNorm}\left(x + \text{Sublayer}(x)\right)$, 
where $\text{Sublayer}(x)$ is the function implemented by either 
the self-attention or the position-wise feed-forward network. 
The residual connection\cite{he2015a} makes it simple for the 
architecture to perform the identity operation on the input $x$ 
since $\text{Sublayer}(x)$ can easily be trained to output 
zeros. The layer normalization\cite{ba2016} is a technique 
to normalize the of intermediate outcome of the sub-layers 
to have zero mean and unit variance, which enables a more 
stable calculation of the gradients in Eq.\ref{eq:gradient}
and faster training.  
 
The embedding mentioned above convert the values of measurements 
$\boldsymbol{a}$ to vectors of dimension $d_\text{model}$ through
a parametrized linear transformation. Since the Transformer 
model contains no recurrence or convolutions, the model can't 
naturally use the information of the the spatial ordering of 
the qubits. To fix this, we include information about the relative 
or absolute position of the measurements in the system by adding 
positional encodings to the input embeddings. The positional 
encodings have the same dimension $d_\text{model}$ as the 
embeddings and are added to the original embedding.\cite{AttentionIsAllYouNeed}  
The last element of the Transformer is a linear layer followed 
by a softmax activation that outputs the conditional distribution.

While several Transformer layers of the type encircled in the 
orange block in Fig.~\ref{fig:transformer} can be composed to 
enhanced the expressiveness of our models, we found that the 
one-layer Transformer-encoder architecture described above suffices
for modeling distributions we are interested in. We train 
the models using Adam Optimizer\cite{Adam} in Pytorch~\cite{pytorch} 
with an initial learning rate of 0.01. Our implementation uses
single-precision (32-bit) floating-point representation for real numbers
and a 64-bit representation for integers. 

\noindent \textbf{Experiments and results}: We demonstrate our approach 
on three types of quantum circuits: circuits producing GHZ states, 
circuits producing one-dimensional graph states, and a variational circuit 
for the ground state of the Transverse Ising Field Model at 
the critical point \cite{vqe2019} (see Appendix). We use a variety 
of quality metrics to quantify the efficacy of our method: 
The KL divergence of Eq.~\ref{eq:KL}, the classical fidelity 
$F_c(\boldsymbol{P}_1,\boldsymbol{P}_2)=\sum_{\boldsymbol{a}} \sqrt{P_{1}(\boldsymbol{a}) 
P_2(\boldsymbol{a})}$, and the $L_1$ norm of the probability distributions 
are all designed to measure the difference between the probability 
distribution of the neural probabilistic model and the exact probability 
distribution (either $\boldsymbol{P}_{i+1}$ or $\boldsymbol{P}_{i+1}^{(e)}$) 
of the POVM.  Note that these measures directly bound how far off the 
POVM measurement statistics of the actual quantum state differ from 
our simulation. These measures depend on the POVM basis; we can also 
directly compare basis-independent quantities such as the `quantum fidelity' 
of the state, 
\begin{equation}
F(\varrho_1, \varrho_2)=\text{Tr}\left[\sqrt{\sqrt{\varrho_1}\,\varrho_2\,\sqrt{\varrho_1}}\right]
\end{equation}  
and 
\begin{equation}
F_2(\varrho_1,\varrho_2)= \sqrt{1- ||\varrho_1-\varrho_2||^2_2/2}\, \lo{.}
\end{equation}
Both $F$ and $F_2$ are equal to the overlap $|\langle \Psi_1 | 
\Psi_2\rangle|$ when $\varrho_1$ and $\varrho_2$ represent pure 
states. Notice that $F_2$ is well defined even if the `density matrices' 
generated from the POVM probability don't correspond to physical density 
matrices.

\begin{figure*}
\centering
\includegraphics[width=\textwidth]{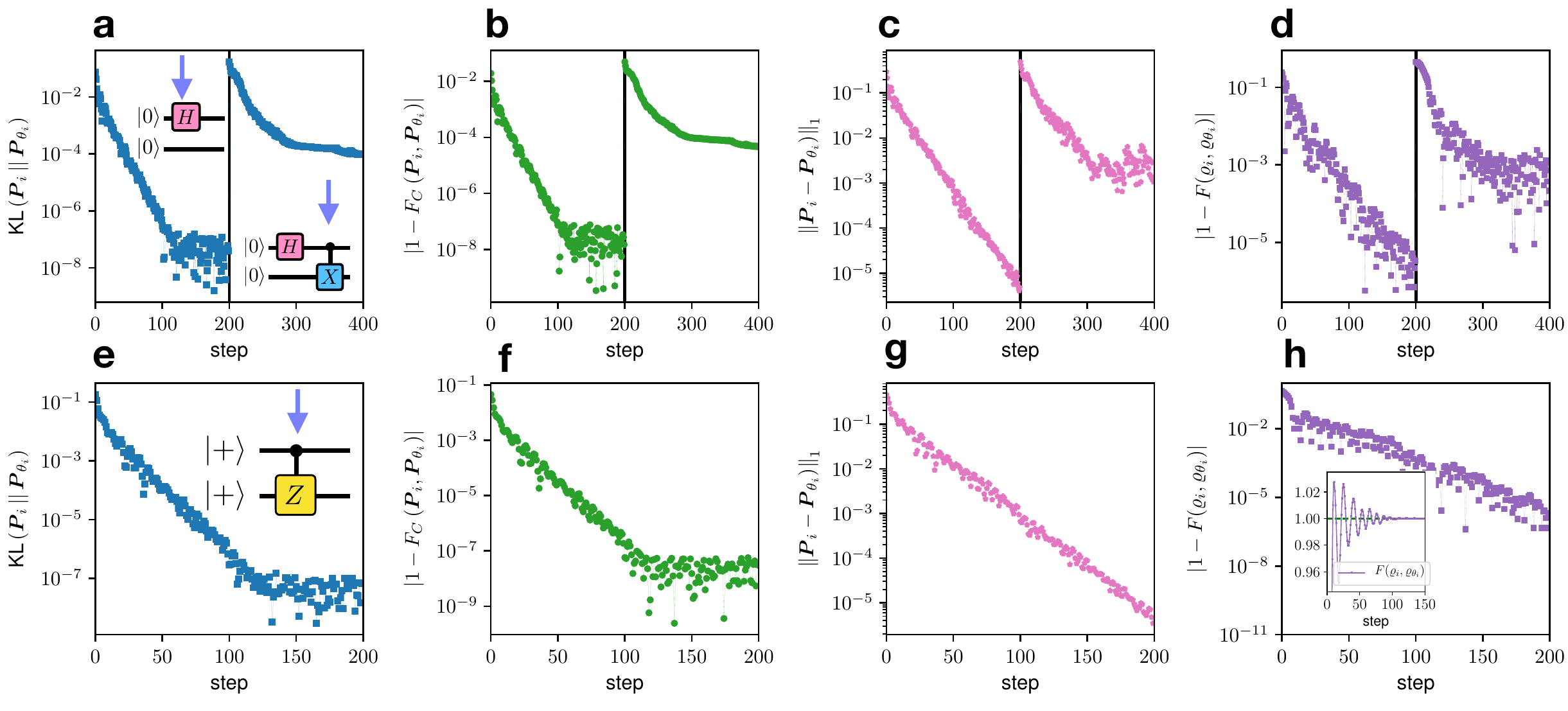}\newline
\caption{Measures of training of two qubit circuits (shown in insets of (a) and (e)) 
between the exact state and the quantum state represented by a Transformer 
($d_\text{model}=16$) after each circuit element. KL divergence (a and e), 
classical fidelity (b and f), $L_1$ norm (c and g) and the quantum fidelity 
(d and h). The main panels use a log-linear scale whereas the inset in (h) 
displays the fidelity in linear scale.}
\label{fig:training}
\end{figure*}

Fig.\ref{fig:training}(a-d) shows these measures for two quantum gates 
(see Fig.\ref{fig:training}(a)[inset]) which generate the GHZ state with 
$N=2$ qubits, namely the  Bell state 
$|\Psi\rangle=\frac{1}{\sqrt{2}}\left(|00\rangle +|11\rangle \right)$. 
For the application of each of the two gates, the KL divergence, the 
L1 classical error, the classical fidelity error, and the quantum fidelity 
error all initially approach zero exponentially in the number of steps. 
The quantum fidelity oscillates around one until finally settling at 1 
by the end of the optimization. The KL divergence and the classical 
fidelity error both eventually saturate, but, interestingly, the L1 
classical error and the quantum fidelity error both continue to improve 
for the application of the first gate in the circuit. This suggests 
that further improvements due to better training of the ansatz and a 
better choice of objective function are possible. The observed
saturation at $\sim 10^{-8}$ also suggests some quantities are limited
by the 32-bit floating-point precision used in computations. Increasing
the precision could also lead to improved convergence. In 
Fig.~\ref{fig:training}(e-h), we present analogous results for a 
circuit generating a $2$-qubit graph state, i.e. 
$|\Psi\rangle=\frac{1}{2}\left(|00\rangle + |10\rangle +|01\rangle -|11\rangle \right)$, 
where we observe similar behaviour.  Note that for these examples, 
we are primarily probing the quality of our optimization given 
that the Transformer with hidden dimension 16 should be powerful 
enough to exactly represent the exact probability distribution. 

The small oscillations of the fidelity above 1.0 evident in the 
inset of Fig.~\ref{fig:training}(h) exists because the Transformer 
can represent probability distributions without a corresponding 
physical density matrix. This is because only a subset of the 
probability simplex, which is the space where the distributions 
we are interested in live, corresponds to physical density 
matrices upon inversion in Eq.\ref{Eq:rho}. The subset of 
probability distributions with a valid quantum states in our 
setting forms a convex set similar to the so-called {\it Qplex} 
in quantum Bayesian theory.\cite{appleby2017} Here we emphasize 
that the fidelity values in Fig.\ref{fig:training}(d) and (h) 
eventually converge to one and the oscillations above and below 
1 are suppressed exponentially with the training steps, 
suggesting that the model converges to the target quantum 
state. We provide more details about the Qplex and the presence 
of unphysical states in our representation in the Appendix. 

\begin{figure}
\centering
\hspace{-0.1cm}\includegraphics[width=0.48\textwidth]{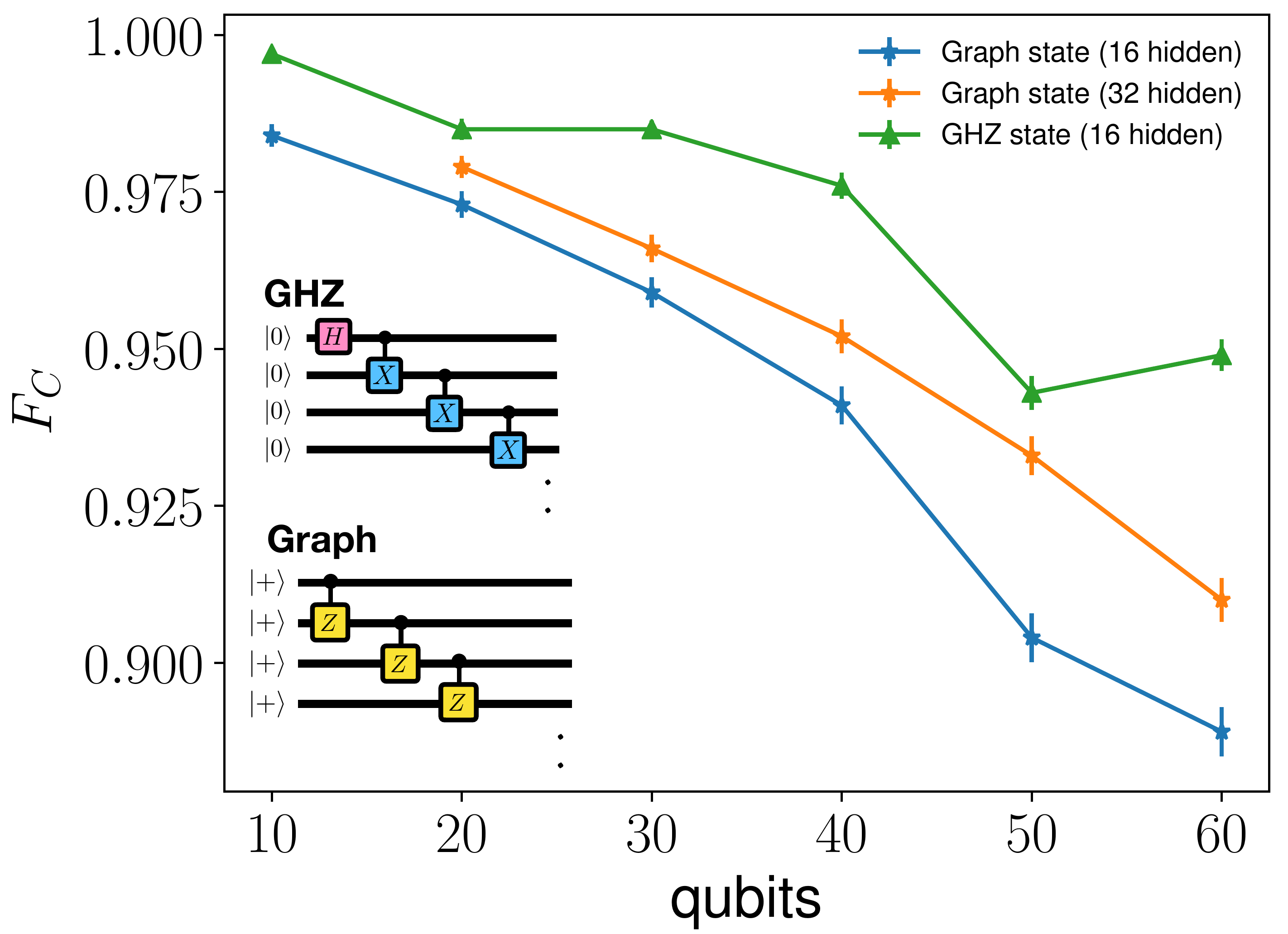}
\caption{Classical fidelity $F_c$ between the final probability distribution of 
the Transformer versus the exact POVM measurements of the post-circuit quantum 
states as a function of the total number of qubits for circuits (see insets) 
generating the GHZ state and linear graph states. The latter is shown with 
Transformers of hidden dimensions 16 and 32. }
\label{fig:multiqubits}
\end{figure}

We now turn our attention to the quality of the circuit 
simulation as a function of the number of qubits in the circuit, 
letting both the depth and gate number grow linearly with the 
number of qubits. We find in constructing GHZ and linear graph 
states (see Fig.~\ref{fig:multiqubits}(inset)) in the range from 
10-60 qubits that the classical fidelity falls approximately linearly 
with number of qubits (see Fig.~\ref{fig:multiqubits}) reaching 
a classical fidelity of approximately 0.9 at 60 qubits. In addition, 
we have considered two different hidden dimensions, 16 and 32, and 
find that there is an improvement of the classical fidelity 
over all qubit sizes as we increase the hidden dimension.  We attribute 
this to an improved representability power of the larger Transformers 
suggesting that one of the bottlenecks of our simulation is the 
ability of our neural probabilistic model to represent the probability 
distribution that corresponds to the output of the quantum circuit.   

We further apply our method for state preparation of a 6-qubit TFIM ground 
state using a variational quantum eigensolver circuit \cite{vqe2019} 
(see Appendix). The variational quantum eigensolver~\cite{peruzzo2014b} 
(VQE) is a quantum/classical hybrid algorithm that can be used to approximate
the lowest energy eigenvalues and eigenvectors of a qubit Hamiltonian
$H$ on a quantum processor. Rather than performing an optimization
of the VQE ansatz, we focus on the probabilistic preparation of an
already optimized VQE circuit for the ground state of the TFIM, as
demonstrated below. We note that the particular circuit we consider 
has more gates per qubit than our previous examples. However, we 
limit our simulation to a small number of qubits so that the 
estimation of quantum fidelity, whose computational cost is exponential 
in the number of qubits in our approach,\cite{carrasquilla2019a} 
remains possible.  
Thus we evaluate both classical and quantum fidelity 
after the application of each quasi-stochastic gate in the circuit 
(see Appendix F for a precise specification of the quantum circuit 
and details of its probabilistic preparation). Both the classical 
and quantum fidelities shown in Fig.~\ref{fig:vqe} drop roughly 
linearly with the number of gates in the circuit; in fact, as 
demonstrated in the Appendix in Fig.~\ref{fig:correlation}, there 
is a clear correlation between the classical and quantum fidelity. 
It is natural to expect that this linear drop observed in our 
simulation is brought on by an accumulation of errors building up 
after successive gates in the circuit. We can give further evidence 
of this by looking at the error made after a single step, 
$F_2(\varrho_i,\varrho_i^{(e)})$, which directly compares 
$\boldsymbol{P}_i^{(e)}$  and $\boldsymbol{P}_{\theta_i}$  
(see Fig.~\ref{fig:vqe}). We find that the single-step error is 
roughly constant and small throughout the circuit suggesting each 
step of the simulation is fairly accurate. This is consistent 
with the observations in Fig.~\ref{fig:training}.   

\begin{figure}
\centering
\includegraphics[width=0.48\textwidth]{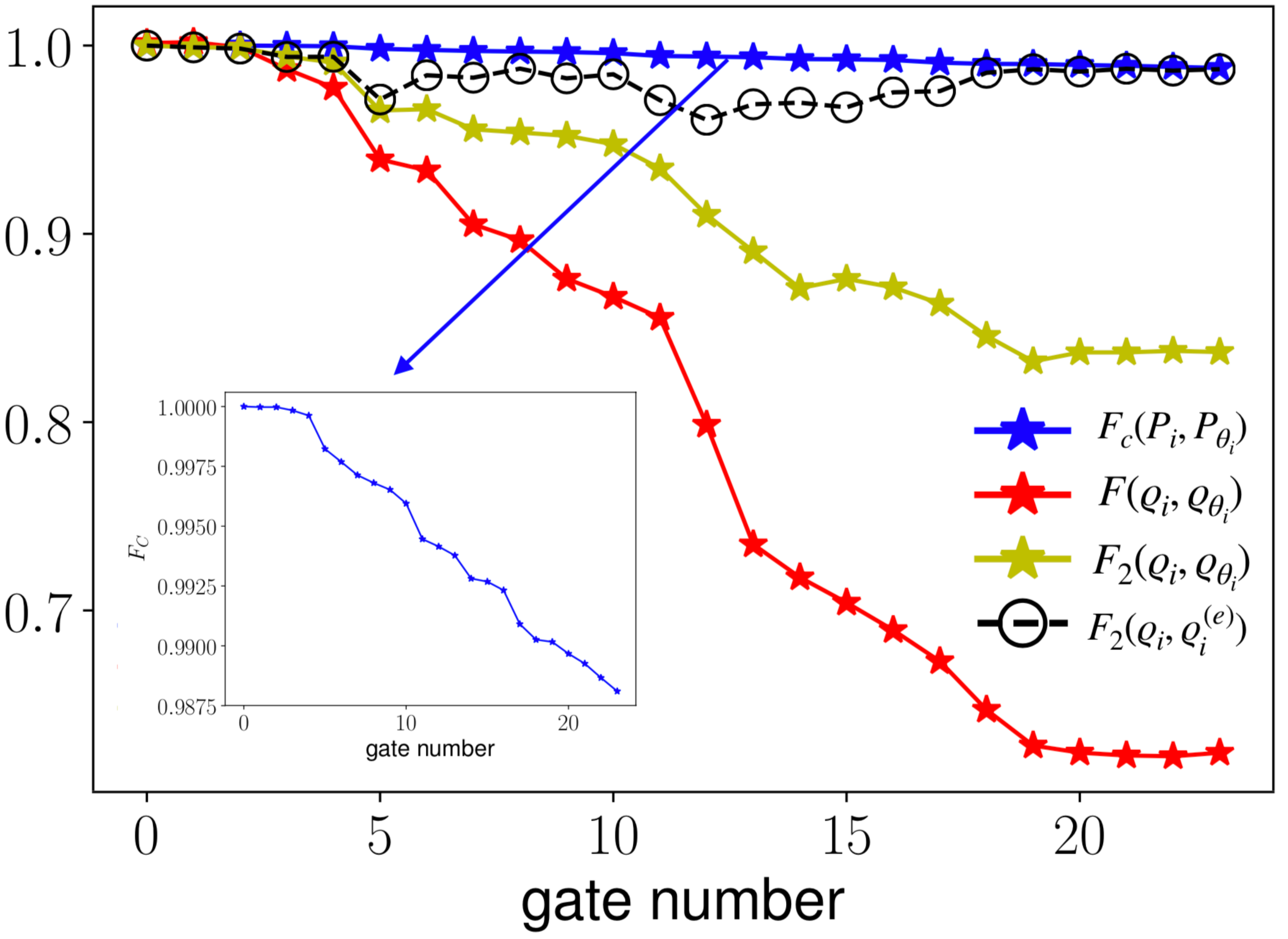}
\caption{Fidelity measurements of a VQE circuit for the preparation of a 
6-qubit ground state of the TFIM as a function of the gate number. The inset 
shows the classical fidelity which is seen to decrease approximately 
linearly with the number of gates applied to the state.}
\label{fig:vqe}
\end{figure}

\noindent \textbf{Conclusions}: We have introduced an approach for the classical 
simulation of quantum circuits using probabilistic models and validate it on a 
number of different circuits.  This is done by using a POVM formalism which maps 
states to probability distributions and gates to quasi-stochastic matrices.  
To represent the probability distribution over this POVM basis, we use a  
Transformer, a powerful machine learning architecture heavily used in natural 
language processing. We develop an efficient sampling scheme which updates 
the Transformer after each gate application within the quantum circuit.  This 
sampling scheme works well out to a large number of qubits; in this work 
we demonstrate simulations up to 60 qubits and empirically see that the accuracy 
of the simulation drops roughly linearly with the number of qubits at a fixed 
hidden dimension of the Transformer architecture. We observe that increasing 
the hidden dimension of the model improves our results for the circuits we 
considered. Although this observation suggests that our approach is scalable, 
a detailed study of the representational power of the Transformer and its 
relation to optimization over a wider class of circuits and Transformer 
hyperparameters is required to properly establish the scalability and 
applicability of our approach. 

Optimizing the Transformer after each gate is a critical step in our  algorithm.  
While already reasonably efficient, there are various ways this optimization might 
be further improved. For example, our current simulations allow probabilistic representations 
which don't map to physical quantum states (i.e. outside of the Qplex). We anticipate that 
constraining the optimization to the physically relevant subspace would improve the quality
of the simulations and their broad applicability. Additionally, further strategies from 
machine learning may be applicable; a common training strategy in natural language 
processing has been to simultaneously train multiple models selecting the best one at 
each step and this technique may improve the accuracy in our quantum circuit simulations.

The choice of the POVM basis directly affects the structure of the underlying probability 
distributions describing the quantum states as well as the efficiency of their simulation. 
Here we chose a simple IC-POVM basis which is single-qubit factorable, which means that 
all of the entanglement and complexity associated with the quantum state can be traced back 
to $\boldsymbol{P}$ and not the POVM basis. Practically, the factorized IC-POVM
representation ensures local unitaries and quantum channels map to local quasi-stochastic 
matrices allowing for the design of practical algorithms.  A common alternative POVM basis, the SIC-POVM\cite{ferrie2011,fuchs2011a,chruscinski2015,vandewetering2018,kiktenko2019} has an 
elegant formalism but is more difficult to work with algorithmically since SIC-POVM basis are 
not known to exist for large systems \cite{fuchs2017} and don't map local unitaries to local 
matrices.  It is nonetheless an interesting research question whether these more complicated 
basis can be useful in the context of numerical simulations. Indeed, POVM is related to the 
Wigner-function quasi-probability representation and it will also be worth further investigating 
their relation. 

While in this work we have stored the probability distribution using a Transformer, there are 
other options for storing this probability distribution including other machine-learning 
architectures and tensor networks.  In fact, it is not even necessary to explicitly store the 
representation at all;  instead, in the spirit of quantum Monte Carlo, it could be sampled 
stochastically.   While such a simulation will generically have a sign problem, there may be 
preferred basis choices for the POVM which minimize that effect for a particular set of quantum circuits.

In general, the classical simulation of quantum circuits is known to be 
difficult.~\cite{supremacy}  Nonetheless, in the era of noisy intermediate-scale 
quantum technology it is important to be able to benchmark machines which have 
qubit sizes that are outside the limits of what can be simulated exactly on 
classical computers to validate and test quantum computers and algorithms.  
Moreover, the ability to simulate ever larger and more difficult circuits helps 
better delineate the boundary between classical and quantum computation.  The 
number of approaches for simulating quantum circuits is small and our approach 
introduces an alternative to the standard approach of simulating the 
quantum state either explicitly,\cite{deraedt2007,markov2008,smelyanskiy2016,pednault2017,
chen2018ClassicalSimulation,markov2018,li2018,jonsson2018} or 
stochastically.\cite{cerf1998,PhysRevLett.109.230503} We anticipate advantages 
with respect to established algorithms enabled by the ability of Transformers
to model long-range correlations,\cite{shen2019} the autoregressive nature of the 
model, as well as the nature of the self-attention mechanism, which allows a high 
degree of parallelization of most of the computations required 
in our approach. Additionally,  extensions of the model which encode information 
about the spatial structure of the problem (e.g. two-dimensional 
Transformers\cite{parmar2018}) can be easily defined while retaining all 
the computational and modelling advantages of the Transformers used in this work. 

Beyond the simulation of circuits, our POVM approach can be naturally extended to various problems in quantum many-body systems, such as the simulation of real-time dynamics of closed and open quantum systems (see Appendix A and B). Thus our work opens up new possibilities for combining the POVM formalism with different numerical methods, ranging from quantum Monte Carlo to machine learning to tensor networks, in an effort to better classically simulate quantum many-body systems.

\acknowledgements
We would like to thank Martin Ganahl, Jimmy Ba, Amir-massoud 
Farahmand, Andrew Tan, Lei Wang, Jianqiao Zhao, Emily Tyhurst, 
G. Torlai, and R. Melko for useful discussions. We also thank the 
Kavli Institute for Theoretical Physics (KITP) in Santa Barbara and the 
program ``Machine Learning for Quantum Many-Body Physics''. This research 
was supported in part by the National Science Foundation under Grant 
No. NSF PHY-1748958. This work utilizes resources supported by the National Science Foundation's Major Research Instrumentation program, grant \#1725729, as well as the University of Illinois at Urbana-Champaign. J.C. acknowledges support from the Natural Sciences 
and Engineering Research Council of Canada (NSERC), the Shared Hierarchical 
Academic Research Computing Network (SHARCNET), Compute Canada, and the 
Canada CIFAR AI chair program. BKC acknowledges support from the Department 
of Energy grant DOE de-sc0020165.
L.A. acknowledges financial support from 
the Brazilian agencies CNPq (PQ grant No. 311416/2015-2 and INCT-IQ), 
FAPERJ (JCN E-26/202.701/2018), CAPES (PROCAD2013),
and the Serrapilheira Institute (grant number Serra-1709-17173).
Research at Perimeter Institute is supported by the Government of Canada 
through the Department of Innovation, Science and Economic Development 
Canada and by the Province of Ontario through the Ministry of Research, 
Innovation and Science.

\appendix

\newcommand{\beginsupplement}{%
        \setcounter{table}{0}
        \renewcommand{\thetable}{A\arabic{table}}%
        \setcounter{figure}{0}
        \renewcommand{\thefigure}{A\arabic{figure}}%
}
\beginsupplement


\section{Quantum channels}

A complete description of the evolution of closed and open 
quantum systems is fundamental to the understanding and 
manipulation of quantum information devices. In contrast 
to closed quantum systems, the evolution of open quantum 
systems is not unitary. Instead, the evolution of the 
density operator of an open quantum system is described 
by the action of a quantum operation or quantum channel, 
which is specified by a completely positive trace-preserving 
(CPTP) maps between spaces of operators.~\cite{Nielsen}
A commonly used representations of CPTP-maps is the 
Kraus or operator-sum  representation\cite{kraus1983} 
where a CPTP-map $\mathcal{E}$ acts on a quantum state 
$\varrho$ as 
\begin{equation}
    \mathcal{E}\left(\varrho\right) = \sum_{\alpha}^{D} K^{(\alpha)} \varrho K^{(\alpha)\dag}, 
\end{equation}
where $\sum_{\alpha}^{D}K^{(\alpha)} K^{(\alpha)\dag}=\openone$. 
The set of matrices $\{ K^{(\alpha)}, \,\,\alpha=1,\dots,D  \}$ 
act on the Hilbert space of the qubits and can be thought of 
as an array with 3 indices. The maximum value of $D=4^N$, 
and the minimum is $D=1$, which corresponds to a unitary transformation.  

Similar to the unitary evolution, if the initial quantum state 
$\varrho$ is prescribed in terms of the outcome statistics 
of an IC-POVM $\boldsymbol{P}$, we can track its evolution 
under a CPTP-map directly in its probabilistic 
representation:
\begin{equation} \label{eq:stoevolKrauss}
 P_{\mathcal{E}}\left(\boldsymbol{a}'' \right) = 
\sum_{\alpha}^{D}\text{Tr}\left[K^{(\alpha)} \varrho 
K^{(\alpha)\dag} M^{(\boldsymbol{a}'')} \right] = 
\sum_{\boldsymbol{a}'} O_{\boldsymbol{a}''\boldsymbol{a}'} 
P(\boldsymbol{a}')
 \end{equation}
where  
\begin{equation}\label{SWStochKrauss}
O_{\boldsymbol{a}''\boldsymbol{a}'}=\sum_{\boldsymbol{a},\alpha} \,
\text{Tr}\left[K^{(\alpha)} M^{(\boldsymbol{a})} K^{(\alpha)\dag} 
M^{(\boldsymbol{a}'')} \right] \,T^{-1}_{\boldsymbol{a},\boldsymbol{a}'}
\end{equation}
is a quasi-stochastic matrix since, as in the unitary case, the values 
in each column add up to 1 but its entries can be positive or 
negative.\cite{curgus2007,curgus2015,chruscinski2015,vandewetering2018}
 
\begin{figure}
    \centering
    \includegraphics[width=0.35\textwidth]{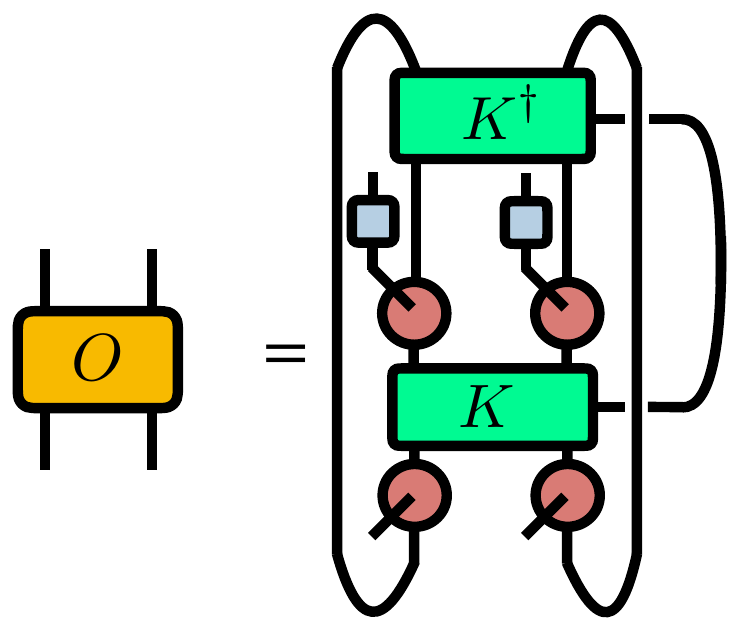}
    \caption{Tensor network representation of $O_{\boldsymbol{a}''\boldsymbol{a}'}$ corresponding
    to a quantum channel that acts on two qubits. The green tensors represent the Kraus 
    operators, which specify the quantum channel and are understood as a rank-3 array.}
    \label{fig:TNrepKRAUSS}
\end{figure}

If a quantum channel acts nontrivially on only $k$ qubits, it 
implies that the quasi-stochastic matrix $O_{\boldsymbol{a}''\boldsymbol{a}'}$ 
acts also only on $k$ qubits. The relation between the quasi-stochastic 
gates and the local quantum channel in Eq.\ref{SWStochKrauss} is graphically 
depicted in Fig.\ref{fig:TNrepKRAUSS} using tensor diagrams.

\section{Liouville-Von Neumann Equation in POVM Formulation}
In Eq.\ref{eq:stoevol} we have discussed how unitary evolution on a 
quantum state in the traditional density matrix formulation translates 
into the factorized IC-POVM formulation used in our study. Accordingly, 
the unitary dynamics induced by Hamiltonian $\mathcal{H}$ acting on 
the system during an infinitesimal time $\Delta t$, i.e., 
$U(t)=e^{-i \Delta t \mathcal{H}}$, implies an equation of motion 
for the measurement statistics
\begin{equation}
\label{swLiou}
i\frac{\partial P\left(\boldsymbol{a}'', t \right)}{\partial t}=
\sum_{\boldsymbol{a},\boldsymbol{a}'} \text{Tr} 
\left(\left[ \mathcal{H},M^{(\boldsymbol{a})} \right] 
M^{(\boldsymbol{a}'')}\right) 
T^{-1}_{\boldsymbol{a},\boldsymbol{a}'}P\left(\boldsymbol{a}', t \right).
\end{equation}
This is equivalent to the Liouville-Von Neumann equation 
$i \frac{\partial \rho}{\partial t}=\left[\mathcal{H},\rho\right]$. 

A solution to Eqs.\eqref{swLiou} for a time-independet Hamiltonian 
is given by $\boldsymbol{P}(t)= e^{-iAt}\boldsymbol{P}(0)$, where 
the matrix elements $A_{\boldsymbol{a}''\boldsymbol{a}'}=\sum_{\boldsymbol{a}} 
T^{-1}_{\boldsymbol{a},\boldsymbol{a}'}\left[ \text{Tr} 
\left(\left[ \mathcal{H},M^{(\boldsymbol{a})}\right] M^{(\boldsymbol{a}'')}\right) \right]$. 

\section{Linblad Equation in POVM Formulation}
Applicable to open quantum systems, an infinitesimal Markovian but 
non-unitary evolution leads to the equivalent of the Linblad equation
\begin{equation}
i\frac{\partial P\left(\boldsymbol{a}, t \right)}{\partial t}= 
\sum_{\boldsymbol{a}} A_{\boldsymbol{a}'',\boldsymbol{a}'} P\left(\boldsymbol{a}, t \right), 
\end{equation}
where the matrix elements are augmented to 
\begin{equation}
\label{swLinb}
\begin{split}
A_{\boldsymbol{a}''\boldsymbol{a}'} & = \sum_{\boldsymbol{a}} T^{-1}_{\boldsymbol{a},\boldsymbol{a}'}\left( \text{Tr} \left(\left[ \mathcal{H},M^{(\boldsymbol{a})}\right] M^{(\boldsymbol{a}'')}\right) \right. \\
& + \sum_k \left[  -\frac{i}{2}\text{Tr}\left( \{ L^{\dag}_k L_k, M^{(\boldsymbol{a})} \} M^{(\boldsymbol{a}'')} \right)  \right. \\
& + \left. \left. i \,\text{Tr}\left( L_kM^{(\boldsymbol{a})} L^{\dag}_k M^{(\boldsymbol{a}'')} \right)\right] \right) 
\end{split}
\end{equation}
Here, the operators $L_k$ are called Lindblad operators or quantum jump 
operators. Like the Liouville-Von Neumann equation, 
Eq.\eqref{swLinb} has a solution for a time-independet Hamiltonians 
given by $\boldsymbol{P}(t)= e^{-iAt}\boldsymbol{P}(0)$.

\section{Measurements}
Although the probabilistic representation of the quantum state 
in Eq.\ref{Eq:rho} already provides the measurement statistics 
of the factorized POVM $M^{(\boldsymbol{a})}$, the statistics 
of other POVMs $\Pi^{(\boldsymbol{b})}$, e.g. a POVM describing 
standard measurements in the computational basis and other 
experimentally relevant operators, are related to 
$M^{(\boldsymbol{a})}$ via the Born rule: 
\begin{equation*}
\begin{split}
P_{\Pi} (\boldsymbol{b})&=\sum_{\boldsymbol{a},\boldsymbol{a}'} P(\boldsymbol{a}')\,
T^{-1}_{\boldsymbol{a},\boldsymbol{a}'}\,\text{Tr}\left[M^{(\boldsymbol{a})}\Pi^{(\boldsymbol{b})}\right]\\
&= \sum_{\boldsymbol{a}'} q(\boldsymbol{b}| \boldsymbol{a}') P(\boldsymbol{a}')\\
\end{split}
\end{equation*}
where $q(\boldsymbol{b}| \boldsymbol{a}')=\sum_{\boldsymbol{a}} T^{-1}_{\boldsymbol{a},\boldsymbol{a}'}\,
\text{Tr}\left[M^{(\boldsymbol{a})}\Pi^{(\boldsymbol{b})}\right] $ can 
be characterized as a quasi-conditional probability distribution since 
its entries can either be positive or negative but its trace over 
$\boldsymbol{b}$ is the identity $\openone_{\boldsymbol{a}'}$. Due 
to its evocative resemblance with the law of total probability, 
the relation between measurement statistics 
$P(\boldsymbol{b})$ and $P(\boldsymbol{a})$ is often called the 
quantum law of total probability in quantum Bayesianism.\cite{fuchs2013}

\section{Qplex and positivity of quantum states}

\begin{figure}
\centering
\includegraphics[width=\linewidth]{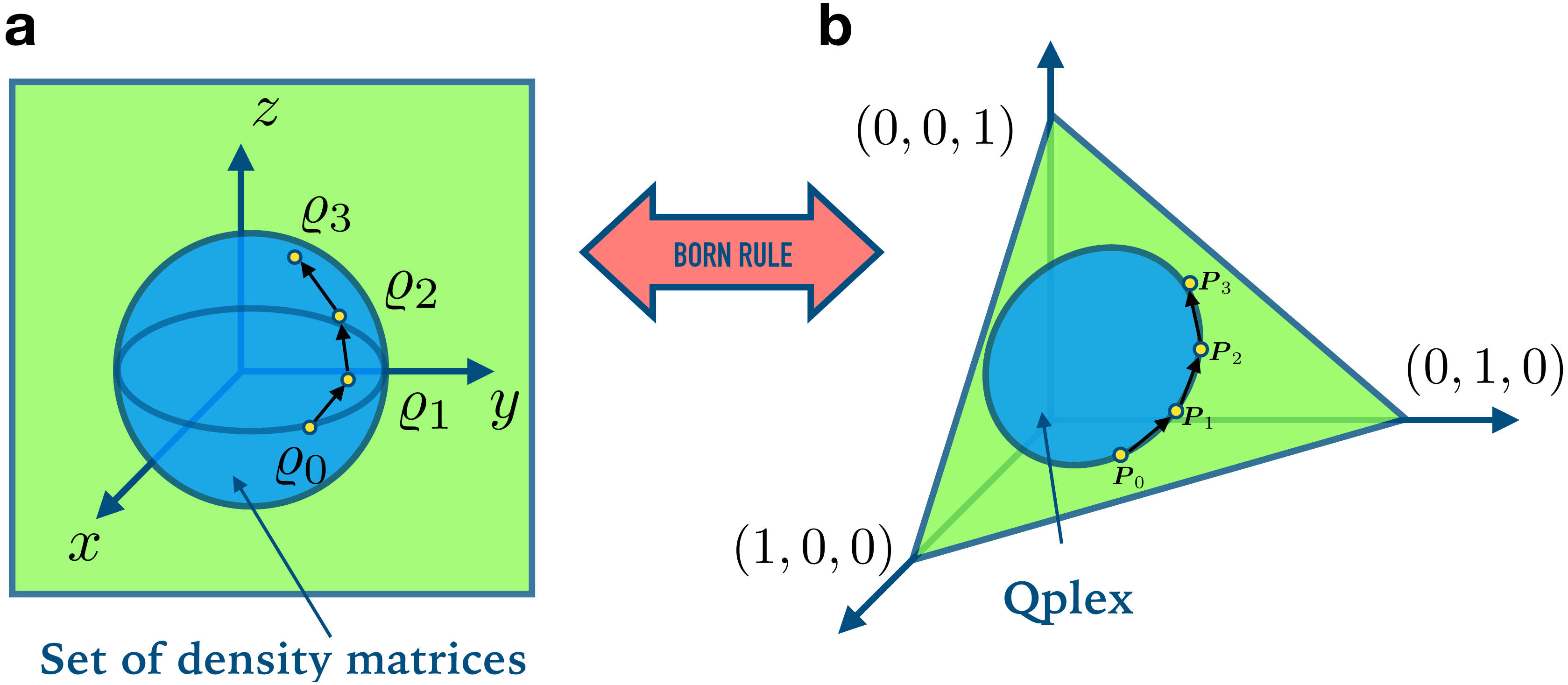}\newline
\caption{Geometry of quantum states. (a) Schematic representation 
of the subset of density matrices (blue sphere). For one qubit, this 
set corresponds to the Bloch sphere. (b) Schematic representation of the 
probability simplex $\Delta_{m^N}$, which represents the set of all 
possible categorical probability distributions with $m^N$ outcomes. A subset 
of these distributions termed Qplex (blue oval) is isomorphic to the 
usual space of quantum states in (a).}

\label{fig:qplex}
\end{figure}

The traditional quantum theory can be viewed as a noncommutative 
generalization of probability theory where quantum states are 
specified by Hermitian, positive semi-definite trace one matrices. 
However, quantum states can also be specified through probability 
distributions corresponding to the statistics of the outcome of 
an informationally complete physical measurement.  From this viewpoint, 
quantum theory is not necessarily a generalization of probability theory; 
instead, it can be seen as augmenting probability theory with further 
rules for dynamics and measurements on quantum systems.\cite{appleby2017} 
When we represent the probabilities of a IC-POVM as points in the 
corresponding probability simplex $\Delta_{4^N}$, these probabilities 
are not arbitrary, since not any point of the simplex $\Delta$ can 
represent a quantum state, only a subset of the simplex. For a
symmetric IC-POVM,\cite{renes2004} this subset is referred to as the 
Qplex.\cite{appleby2017} Even though the IC-POVM used in our work is not 
symmetric, we will still refer to the subset of distributions with a 
corresponding quantum state in Eq.\ref{Eq:rho} as a Qplex. The 
space of all possible states of a given quantum system and the 
corresponding Qplex are schematically represented in 
Fig.\ref{fig:qplex}(a)-(b).  

A small quantum computation in also depicted schematically in 
Fig.\ref{fig:qplex}(a) and (b). This computation starts with a simple 
pure product state $\varrho_0$ followed by the application of three 
unitary matrices which take the state from $\varrho_0$ to $\varrho_3$. 
These computations occur at the boundary separating valid quantum 
states from other operators; such boundary includes all the pure states. 
Correspondingly, since the relation between the space of quantum states 
and the Qplex is linear (i.e. the Born rule), quantum computations 
in the probabilistic language take place at the edges of the Qplex, 
as illustrated in Fig.\ref{fig:qplex}(b). 

While these observations do not have any major conceptual implication 
for the physical realization of quantum computations, this geometric 
interpretation can help us clarify some aspects we observe in the results 
from our simulation strategy. The most important aspect is the fact 
that the probabilistic model in our study, in general, lives in an 
standard simplex $\Delta_{m^N}$ and is not constrained to the subset 
of valid ``quantum'' distributions. Even though the update rule in 
Eq.\ref{eq:stoevol} should produce distributions that live on the Qplex, 
since we use an approximate update, it is possible that the model
may temporarily leave the Qplex. This is observed in 
Fig.\ref{fig:training}(d) and (h) where we observe values of quantum 
fidelity higher than 1, which means that during the training process, 
the ransformer induces matrices in Eq.\ref{Eq:rho} that are not valid 
quantum states. Note that the fidelity in Fig.\ref{fig:training}(d) 
and (h) eventually converges and stays at values very close to one 
and that the oscillations above 1 disappear, suggesting that the
state is getting closer and closer to the target, valid quantum state. 

\section{Variational Circuit for TFIM}

\begin{figure}[H]
\centering
\includegraphics[scale=0.25]{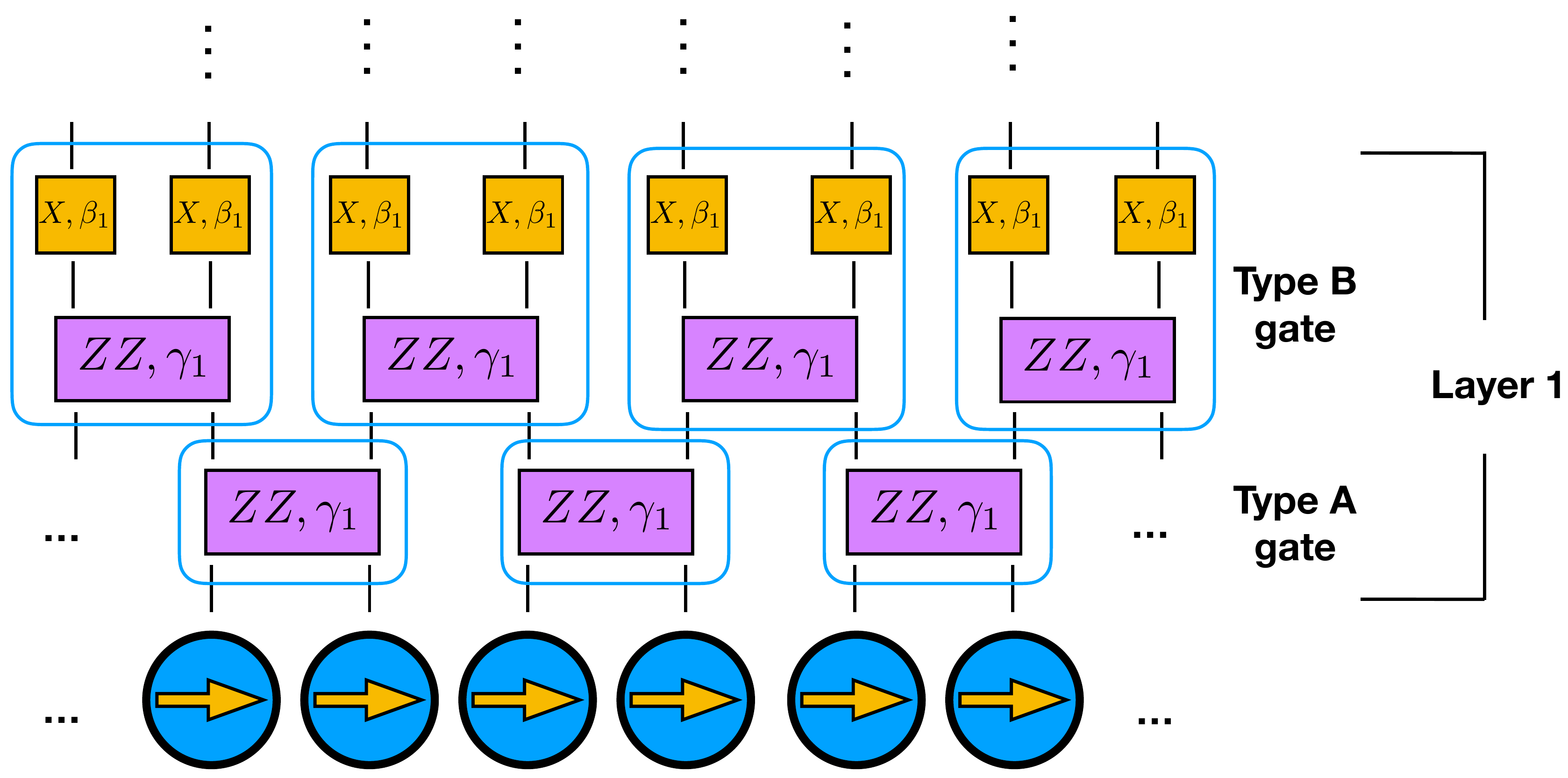}
\caption{Variational circuit preparation for TFIM. Only the first of the 4 layers in our
calculation is shown. The gates encircled in rounded blue squares are combined 
and subsequently transformed into quasi-stochastic gates for the 
probabilistic simulation of the quantum circuit.}
\label{fig:circuit}
\end{figure}

We use the variational circuit depicted in~Fig.\ref{fig:circuit} for 
the 6-qubit TFIM preparation.\cite{vqe2019} The parameters for gamma and beta are 
taken alternatively from the following sequence describing a circuit with 
4 layers, $(0.2496, 0.6845, 0.4808, 0.6559, 0.5260, 0.6048, 0.4503, 0.3180)$. 
Note that in our simulations, we do not directly transform the original gates
into quasi-stochastic gates. To save computational resources, we combine 
the quantum gates encircled in rounded blue squares, after which
we transform them into quasi-stochastic matrices.  

\begin{figure}[H]
 \centering
 \includegraphics[scale=0.5]{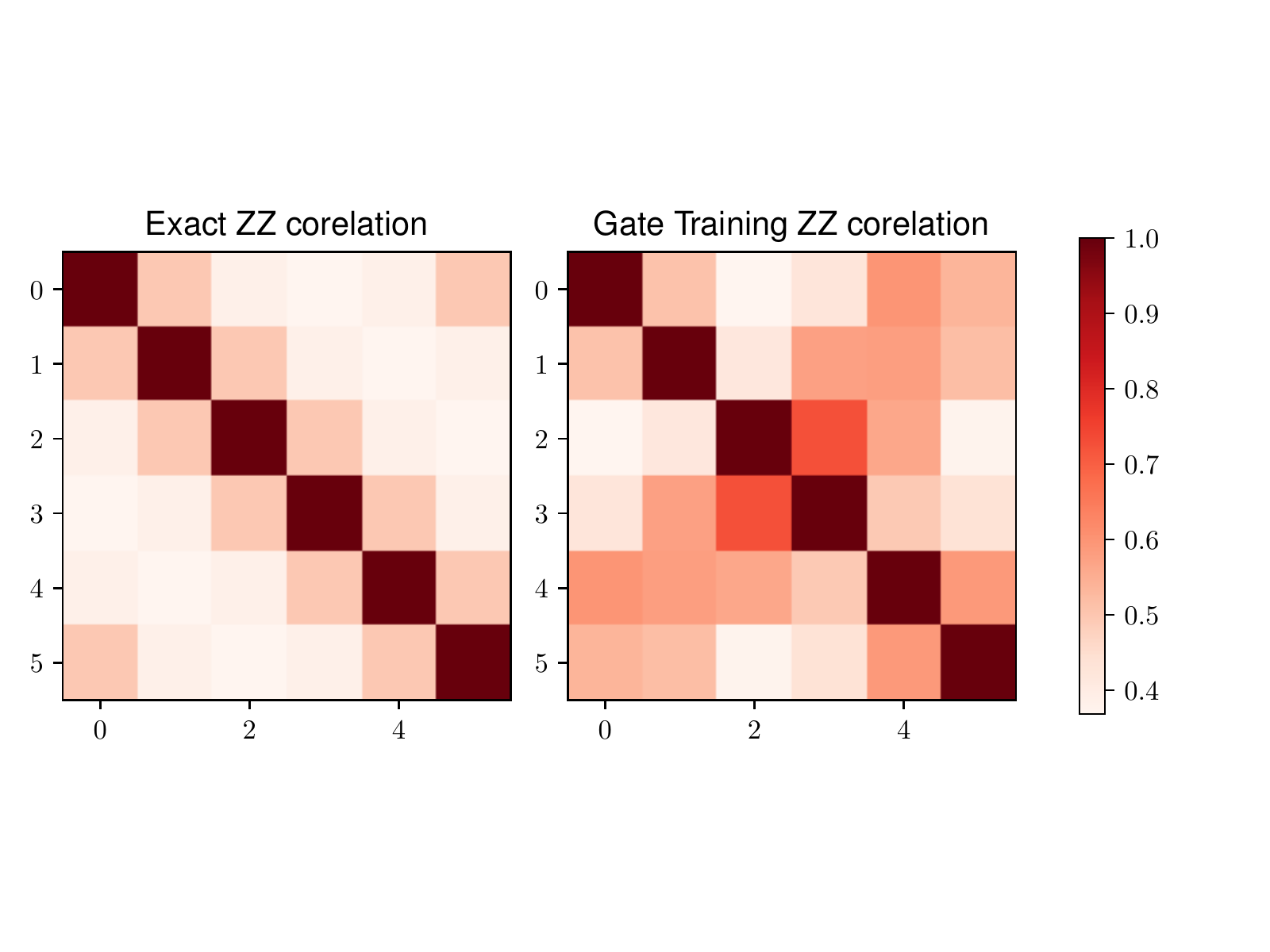}
 \caption{Comparison between $\sigma^{z}_i \sigma^{z}_j$ correlation from exact quantum circuit
  state and the gate training state.}
 \label{fig:ZZ}
\end{figure}

Using this circuit, we have also computed the $\sigma^{z}_i \sigma^{z}_j$ correlation 
of the exact variational circuit in Fig.\ref{fig:circuit} and the POVM trained circuit, 
which are compared in Fig. ~\ref{fig:ZZ}.

\section{Supplementary Figures}

\begin{figure}[H]
\centering
\includegraphics[scale=0.5]{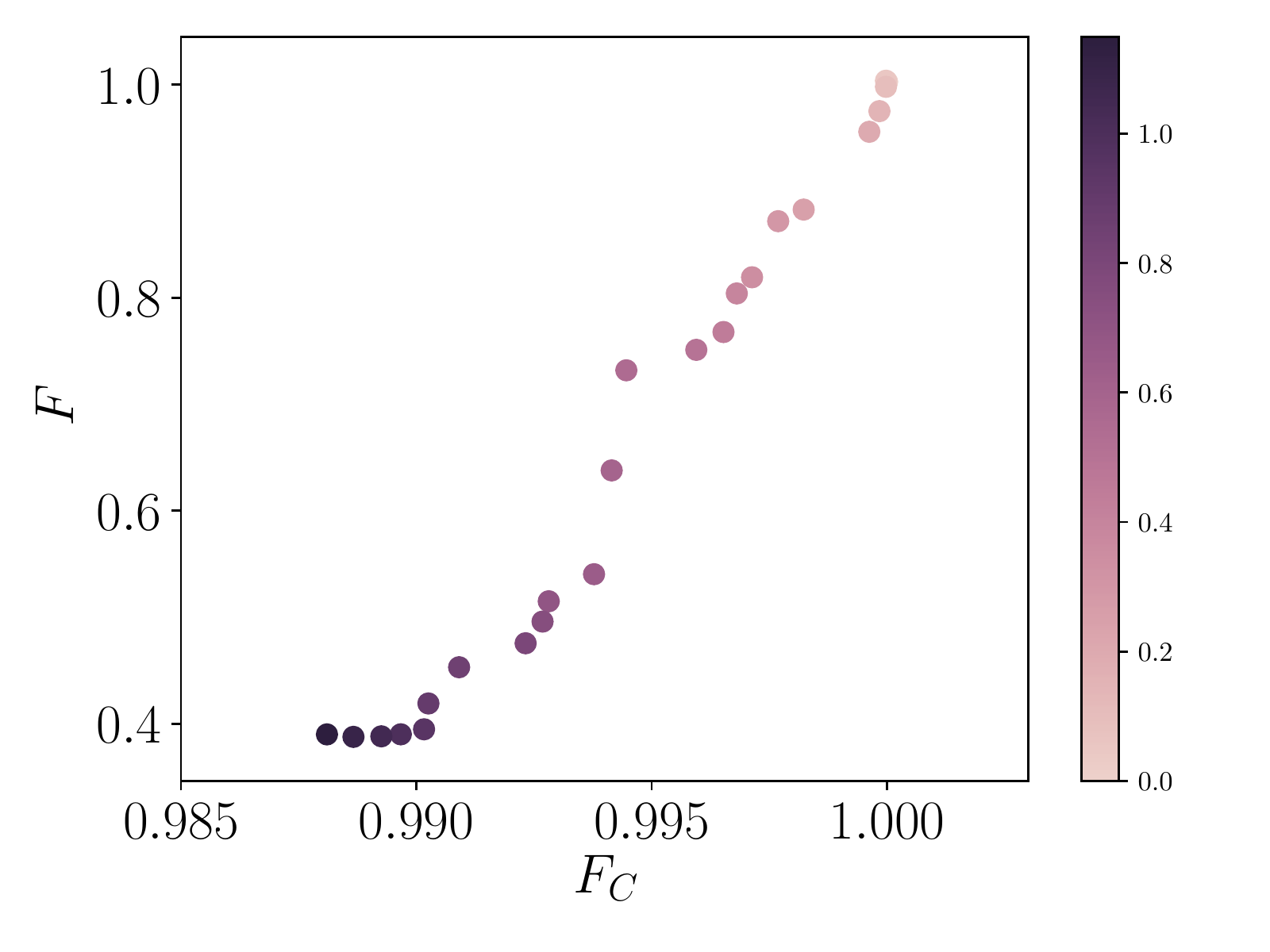}
\caption{Correlation between classical fidelity and quantum fidelity. 
The darker color corresponds to gate that is applied later.}
\label{fig:correlation}
\end{figure}

%

\end{document}